\documentclass[a4paper, amsfonts, amssymb, amsmath, prd, showkeys, nofootinbib, twocolumn, superscriptaddress, colorlinks, linkcolor=refcol, citecolor=refcol, urlcolor=refcol]{revtex4-2}
\usepackage[english]{babel}
\usepackage[utf8]{inputenc}
\usepackage{orcidlink}    
\usepackage[colorinlistoftodos, color=green!40, prependcaption]{todonotes}
\usepackage{bbold}
\usepackage{physics}
\usepackage{makecell}
\definecolor{refcol}{RGB}{178,34,34}
\usepackage{microtype}
\graphicspath{{./figures/}}
\definecolor{red}{rgb}{1,0,0}
 
\definecolor{internationalkleinblue}{rgb}{0.0, 0.18, 0.65}
 
\def\eq#1{(\ref{#1})}

\def\Eq#1{Eq.~(\ref{#1})}

\def\Tab#1{Tab.~\ref{#1}}

\allowdisplaybreaks

\begin{document}
\title{Anomalous $U(1)_A$ couplings and the Columbia plot}

\author{Francesco Giacosa \orcidlink{0000-0002-7290-9366}}
\email{fgiacosa@ujk.edu.pl}

    \affiliation{Institute of Physics, Jan Kochanowski University, ulica Uniwersytecka 7, P-25-406 Kielce, Poland}
    \affiliation{Institute for Theoretical Physics, Goethe-University,
Max-von-Laue-Straße 1, D-60438 Frankfurt am Main, Germany}

\author{Győző Kovács \orcidlink{0000-0003-0570-3621}}
\email{kovacs.gyozo@wigner.hun-ren.hu}
\thanks{corresponding author}
    \affiliation{Institute for Particle and Nuclear Physics,
    HUN-REN Wigner Research Centre for Physics, 1121 Budapest, Konkoly–Thege Miklós út 29-33, Hungary}
    \affiliation{Institute of Theoretical Physics, University of Wroclaw, PL-50204 Wrocław, Poland}

\author{Péter~Kovács~\orcidlink{0000-0003-3735-7620}}
\email{kovacs.peter@wigner.hun-ren.hu}
    \affiliation{Institute for Particle and Nuclear Physics, HUN-REN Wigner Research Centre for Physics, 1121 Budapest, Konkoly–Thege Miklós út 29-33, Hungary}

\author{Robert D. Pisarski  \orcidlink{0000-0002-7862-4759}}
\email{pisarski@bnl.gov}
    \affiliation{Department of Physics, Brookhaven National Laboratory, Upton, NY 11973}

\author{Fabian Rennecke \orcidlink{0000-0003-1448-677X}}
\email{fabian.rennecke@theo.physik.uni-giessen.de}
    \affiliation{Institute for Theoretical Physics, Justus Liebig University Giessen, Heinrich-Buff-Ring 16, 35392 Giessen, Germany}
    \affiliation{Helmholtz Research Academy Hesse for FAIR (HFHF), Campus Giessen, Giessen, Germany}

\date{October 10, 2024}

\begin{abstract}
When the quark masses are lighter than those in QCD, the standard lore is that a chiral transition of first order must emerge for three, light flavors. Recently, however, numerical simulations on the lattice suggest that the chiral transition is of second order in the chiral limit. Using an extended linear sigma model in the mean field  approximation, we study the relation between terms which break the anomalous, $U(1)_A$ symmetry and the order of the chiral phase transition, especially how a chiral transition of second order can arise for three, massless flavors. We note that in an (unphysical) region of the ``Columbia" phase diagram, when the strange quark mass is light and negative, corresponding to topological angle $\theta=\pi$, the $CP$ symmetry is spontaneously broken.
\end{abstract}


\hypersetup{
pdftitle={Anomalous U(1)A couplings and the Columbia plot},
pdfauthor={F. Giacosa, Gy. Kovacs, P. Kovacs, R. D. Pisarski, F. Rennecke}
}

\maketitle

\section{Introduction} \label{sec:intro}

Numerical simulations on the lattice have shown that at nonzero temperature, QCD crosses smoothly from a phase of hadronic matter to one where the chiral symmetry is approximately restored \cite{Aoki:2006we,HotQCD:2018pds,Borsanyi:2020fev,Guenther:2022wcr}. In the thermodynamic limit, true phase transitions are expected to occur either at finite density, or for smaller current quark masses. While the latter is unphysical, it can help shed light on the phase diagram of QCD, most notably on how the axial anomaly (the breaking of the classical $U(1)_A$ symmetry by quantum fluctations) manifests itself. The dependence of the order of the chiral phase transition on the masses of light and strange quarks is depicted in the so-called Columbia plot \cite{Brown:1990ev}, where the order is plotted as a function of the quark masses, with $m_{\rm up} = m_{\rm down}$ and $m_{\rm strange}$.

Since chiral symmetry is broken explicitly by finite current quark masses, phase transitions must occur for massless quark flavors. Refs.\ \cite{Pisarski:1983ms,Pisarski:2024esv} noted that for two massless quark flavors the order of the transition is sensitive to the fate of the axial anomaly: if $U(1)_A$ remains broken by the axial anomaly at the critical temperature, $T_\chi$, the transition could be of second order, whereas if $U(1)_A$ is restored at $T_\chi$,  the $\epsilon$-expansion predicts a first order transition induced by fluctuations.  For three massless quark flavors, a first order transition appears unavoidable, regardless of what happens to the $U(1)_A$ symmetry. These predictions are based on a perturbative renormalization group study of a linear sigma model (LSM) and the $\epsilon$ expansion, extrapolating from $4\!-\!\epsilon$ dimensions to three \cite{Pisarski:1980ix,Pisarski:1981hir,Pisarski:1983ms}.

This scenario has been called into question by numerical simulations on the lattice, which have not found {\it any} evidence for a first order transition, even for pions as light as $m_\pi < 50$~MeV \cite{Bazavov:2017xul,Dini:2021hug} using staggered fermions, or $m_\pi < 100$~MeV using Wilson fermions \cite{Kuramashi:2020meg}. By looking at the position of a tricritical point as a function of the number of flavors, $N_f$, Ref.\ \cite{Cuteri:2021ikv} finds that the chiral transition is of second order in the chiral limit for $2 \leq N_f \leq 6${\,}\footnote{The upper bound is uncertain, and could possibly reach up to the start of the conformal window, which is somewhere between $9 \lesssim N_f \lesssim 12$}. Recent results from Dyson-Schwinger equations also find a second-order transition for three massless quark flavors \cite{Bernhardt:2023hpr}. 

{\it If}\, in the chiral limit the appropriate universality class for the chiral phase transition is not
\begin{equation}
  {\cal G}_{\chi} = SU(N_f)_L \times SU(N_f)_R  \label{Eq:G_chi}
\end{equation}
but that where the anomaly is restored,
\begin{equation}
  {\cal G}_{\rm cl} \equiv SU(N_f)_L \times SU(N_f)_R \times U(1)_A \; ,   \label{Eq:G_cl}
\end{equation}
then old \cite{Pisarski:1980ix,Pisarski:1981hir,Pisarski:1983ms} results indicate that the transition in \hbox{$4-\epsilon$ dimensions} is a fluctuation-induced first order transition for $N_f > \sqrt{2}$. This remains valid in $\sim \epsilon^5$ \cite{Calabrese:2004uk} and Monte Carlo simulations of the scalar theory directly in three dimensions  \cite{Sorokin:2021jwf,Sorokin:2022zwh}. Other analyses \cite{Butti:2003nu, Schaefer:2008hk, Resch:2017vjs, Fejos:2022mso} also find a first order transition for three flavors in the chiral limit.

There is conflicting evidence, however. Using the functional renormalization group (FRG) it was found in Refs.\ \cite{Fejos:2021yod,Fejos:2024bgl} that {\it if} the anomaly is restored, so at $T_\chi$ the global symmetry is ${\cal G}_{\rm cl}$ (Eq.~\eqref{Eq:G_cl}), then in three dimensions, there is a stable infrared fixed point - {\it not} seen in $4-\epsilon$ dimensions - for all
$N_f \geq 2$.  
This agrees with analyses using the conformal bootstrap method for both two \cite{Nakayama:2014sba,Nakayama:2016jhq,Henriksson:2020fqi,Kousvos:2022ewl} and three \cite{Kousvos:2022ewl} flavors.
For $N_f =2$, the existence of a stable fixed point has already been reported in Refs.~\cite{Butti:2003nu, Grahl:2014fna}.
Note, however, that the FRG results in Refs.~\cite{Mitter:2013fxa, Resch:2017vjs} indicate that, in the case of two flavors and restored axial anomaly, the chiral phase transition does not seem to be governed by this fixed point for physical parameters, as the actual transition turns out to be of first order.

The upper right hand corner of the Columbia phase diagram is that of heavy quarks.  For infinitely heavy quarks, the phase transition is determined by the $Z(3)$ 1-form symmetry of a $SU(3)$ gauge theory, and is inescapeably first order \cite{Yaffe:1982qf}. Quarks act like a background $Z(3)$ field, and so wash out the first order transition, so in the Columbia phase diagram, a first order deconfining region then ends in a line of deconfining critical endpoints. Effective models \cite{Kashiwa:2012wa} and lattice simulations \cite{Fromm:2011qi} find that the deconfining critical endpoints arise for rather heavy quarks, whose current quark mass is $m_{\rm deconf} \sim 2$~GeV.

For quark masses below $m_{\rm deconf}$,
thermodynamically there is no true
phase transition, but just a crossover for both
deconfinement and chiral symmetry breaking.
This includes the physically relevant case of QCD
\cite{Aoki:2006we,HotQCD:2018pds,Borsanyi:2020fev,Guenther:2022wcr}.

In the Columbia phase diagram, then, an
interesting region is the lower left hand corner, 
where the quarks are light.
This region is controlled by the interactions generated through the axial anomaly \cite{Pisarski:1983ms,Pisarski:2019upw,Pisarski:2024esv}.

For the chiral phase transition, the axial anomaly manifests
itself through quark anti-quark, and thereby mesonic,
correlations which are generated by topologically
nontrivial fluctuations. With $q_L$ and $q_R$ the left and right
handed quark fields, the first of these correlations is $\sim \det (\bar q_L q_R)$.  For $N_f$ flavors, this 't Hooft determinant
involves $2N_f$ quarks and is generated by instantons with
unit topological charge \cite{tHooft:1976rip, tHooft:1976snw}. Similarly, fluctuations with topological charge $Q$
generate $2 N_f  Q$-quark anomalous couplings
$\sim (\det (\bar q_L q_R))^Q$ \cite{Pisarski:2019upw, Rennecke:2020zgb}.

The 't Hooft operator is the anomalous operator with the lowest
mass dimension.  Since for three flavors it is a cubic operator,
the magnitude of this operator at $T_\chi$
determines how strong the putative first order transition
is in the chiral limit.  Similarly, nonzero quark masses
act like a background
field for the chiral transition, and so weaken a first
order chiral transition, until it ends in a chiral critical
end point. 
Given that recent results from the lattice and functional methods indicate that the first order regime is small,
the coefficient of the 't Hooft operator must be
small at $T_\chi$.  If the chiral transition is of second
order in the chiral limit, then there is no first
order region in the lower left hand corner of the Columbia
phase diagram \cite{Pisarski:2024esv}.

If so, this does not necessarily imply that higher order
anomalous operators, with $|Q| > 1$, vanish. 
Ref.\ \cite{Pisarski:2024esv} suggested that
if the first-order chiral region in the Columbia plot is small, then operators with $|Q| > 1$ dominate over those
with $|Q| = 1$.

How exactly the axial anomaly is realized in terms of different anomalous correlations at low energies affects not only
the order of the chiral transition, but the properties of hadronic matter in vacuum. The viability of any scenario
can be tested by computing hadronic masses, decay widths,
and associated amplitudes in vacuum.
A widely used class of such models are LSM, where various hadronic degrees of freedom can effectively be taken into account
\cite{Parganlija:2012fy}. 
It was found that the vacuum properties
are insensitive to whether the anomaly manifests
itself including terms with $|Q|=1$, $|Q| = 2$, or
combinations thereof
\cite{Parganlija:2012fy, Olbrich:2015gln, Divotgey:2016pst, Parganlija:2016yxq}, especially the exploratory results of Ref.\ \cite{Kovacs:2013xca}.

In this work we investigate the effect of different anomalous correlations on the chiral phase transition using the extended LSM (eLSM) put forward in Refs.~\cite{Parganlija:2012fy,Kovacs:2016juc};for a recent review, see \cite{Giacosa:2024epf}. Our
purpose is to shed light on the underlying physical mechanism that leads to a small, or even vanishing first-order region, in the lower-left corner of the Columbia plot. This may be regarded as a first proof of concept for the conjectures made in Ref.\ \cite{Pisarski:2024esv}.

\section{Extended linear sigma model and anomalous correlations} \label{sec:LSM}

The LSM has been extended with vector and axial-vector mesons to be able to describe the $\bar q q$ resonances below 2 GeV in Ref.~\cite{Parganlija:2012fy, Giacosa:2024epf} with three flavors. A detailed parameterization procedure has been implemented (see Sec.~\ref{sec:vacuum}) to properly describe the vacuum meson phenomenology and to fit the model parameters. For this, the curvature meson masses are calculated as the second derivative of the grand potential with respect to the fluctuating mesonic fields, while the decay widths are determined at the tree level. Following Ref.~\cite{Kovacs:2016juc}, the fermionic thermal and vacuum fluctuations are included so that the model is solved at the mean-field level and thus the meson masses include fermionic one-loop corrections. In addition, a Polyakov loop potential is considered to mimic the effect of confinement at finite temperatures, and as a consequence, the Polyakov loop variables also appear in the fermionic thermal contribution. With these approximations, the model describes the thermodynamics and phase transition at finite temperatures in good agreement with the lattice results. The phase diagram at finite chemical potentials, and thus the existence and location of the critical endpoint, can also be studied.  
The eLSM has already been used to study not only the meson phenomenology \cite{Kovacs:2021kas} (including glueballs \cite{Eshraim:2012jv, Janowski:2014ppa, Giacosa:2023fdz, Vereijken:2023jor}), but also various aspects/properties of the phase diagram, e.g. its large-$N_c$ limit \cite{Kovacs:2022zcl} or the implications of finite volume effects \cite{Kovacs:2023kbv}. It has also been used in astrophysical applications as the high-density part of a hybrid model \cite{Kovacs:2021ger, Takatsy:2023xzf}.

The effective Lagrangian shown in App.~\ref{sec:elsm_details} can be divided into three contributions, $\mathcal{L}= \mathcal{L}_{\rm cl} + \mathcal{L}_{\rm esb}+ \mathcal{L}_{\rm qu}$. $\mathcal{L}_{\rm cl}$ obeys the classical symmetry $ \mathcal{G}_{\rm cl}$ (Eq.~\eqref{Eq:G_cl}). 
$\mathcal{L}_{\rm esb}$ contains explicit breaking terms of the chiral symmetry, in particular through the two parameters, $h_N$ and $h_S$, which are proportional to the nonzero current masses of light and strange quarks, respectively ($h_{N,S} \propto m_{n,s}$ with $m_n =(m_u + m_d)/2$). In general, this leads to $\mathcal{G}_{\rm cl} \to SU(2)_V \times U(1)_A$, and only for $h_S = h_N/\sqrt{2} \neq 0$ there is a larger residual $SU(3)_V \times U(1)_A$ symmetry. 
$\mathcal{L}_{\rm qu}$ contains all anomalous contributions where $U(1)_A$ is broken by quantum effects and is thus invariant under the chiral symmetry $\mathcal{G}_{\chi}$ (Eq.~\eqref{Eq:G_chi}).
Note that we do not explicitly write down the additional vector $U(1)_V$ symmetry associated with baryon number conservation, as it is always preserved here.

For a phase transition at nonzero temperature, the relevant operators are those in three dimensions, and include operators up to sixth order in $\Phi \sim \overline{q}_L q_R$ (for the definition, see Eq.~\eqref{Eq:mfields} of App.~\ref{sec:elsm_details}).  Thus in general, there are three anomalous operators which are relevant or marginal:
\begin{align}
    \begin{split} \label{eq:Lqu}
        \mathcal{L}_{\rm qu} &= -\xi_1 \big(\det\Phi + \det\Phi^\dagger\big)\\
        &\quad - \xi_1^{1}\, {\rm tr}\big( \Phi^\dagger\Phi\big) \big(\det\Phi + \det\Phi^\dagger\big)\\
        &\quad -\xi_2 \Big[\big(\det\Phi\big)^2 + \big(\det\Phi^\dagger\big)^2\Big]\,.
    \end{split}
\end{align}
Note that this is the most general anomalous Lagrangian, as terms $\sim \det\Phi \det\Phi^\dagger$ are not anomalous and can be expressed in terms of invariants under $\mathcal{G}_{\rm cl}$\footnote{In fact, $ 6\det\Phi \det\Phi^\dagger \!=\! 6\det(\Phi \Phi^\dagger) \!=\! (\Tr(\Phi \Phi^\dagger))^3 -3\Tr(\Phi \Phi^\dagger)\Tr((\Phi \Phi^\dagger)^2) + 2\Tr((\Phi \Phi^\dagger)^3)$. 
However, the presence of such a term would not qualitatively change our findings about the Columbia plot.}.
The first, cubic term is the mesonic 't Hooft determinant related to fluctuations of field configurations with topological charge $Q=1$ \cite{tHooft:1976snw}. The last, sixth-order term is generated by fluctuations with $Q=2$ \cite{Pisarski:2019upw}. $(\det\Phi)^Q$ is generated by multi-instantons and $(\det\Phi^\dagger)^Q$ by anti--multi-instantons. The second term mixes anomalous and classical invariants and has to be included in general as it is of fifth order in the meson fields and hence relevant. Note that the anomalous terms have a residual $Z(Q N_f)$ symmetry, but this is already part of the center of $SU(N_F)_{L/R}$.

We use a mean-field approximation here, which entails that the anomalous couplings are not renormalized and hence neither depend on temperature nor the quark masses. Without further external input we therefore cannot test the scenario where $U(1)_A$ is restored only at the chiral phase transition. At least one of the anomalous couplings $\xi_1$, $\xi_1^{1}$ or $\xi_2$ has to be nonzero in order to correctly implement the axial anomaly in vacuum. As discussed in Ref.\ \cite{Pisarski:2019upw} this means in particular that the anomaly is encoded in higher-order correlations if $\xi_1$ vanishes. It would in principle be possible to model the temperature- and mass-dependence of these couplings, but we restrict ourselves to a self-consistent setup in this work.

The strategy we follow to investigate the impact of the strength of chiral anomalous correlations on the order of the chiral phase transition is as follows: We fit experimental data on hadronic vacuum parameters for different values of $\xi_i\in \xi_X$ (where $\xi_X$ is any non-zero subset of $\{\xi_1, \xi_1^1, \xi_2\}$). As discussed in the next section, this procedure is not unique, and we pick representative parameter sets with a ``good" $\chi^2$. It turns out that it is possible to find parameter sets which describe the experimental data reasonably well, but with very different values for $\xi_i$. Note that this is already a nontrivial result. We then solve our model with these parameters at finite temperature for different quark masses and extract the order of the chiral phase transition. This yields Columbia plots of models with the same vacuum phenomenology but different $\xi_i$, allowing us to explore how the chiral transition depends in particular on the strength of the lowest-order anomalous correlation.

Finally, we mention that there can be anomalous terms that contain fields other than those included in the current model -- see \cite{Giacosa:2017pos,Giacosa:2023fdz} -- but these terms do not affect the Columbia plot.

\section{Vacuum phenomenology} \label{sec:vacuum}

The model parameters are determined using a $\chi^2$ fit with 29 physical quantities, including meson curvature masses, tree-level decay widths, decay constants for the pions and kaons, and the pseudocritical temperature $T_{\chi}$ at $\mu_q = 0$ (see $\rm Fit^{1,1,1,2}$ in Ref.~\cite{Kovacs:2016juc} for the full list). We consider 14 parameters listed in App.~\ref{sec:elsm_details}, not including those of the axial anomaly. The fit starts from many random points -- usually in the range of $\order{10^4-10^7}$ -- in the parameter space, optimized by the steepest descent method to find a local $\chi^2$ minimum. To compare scenarios with different numbers of parameters, it is convenient to use the reduced $\chi^2_\text{red}=\chi^2/N_\text{dof}$, where $N_\text{dof}$ is the number of degrees of freedom. With a sufficient number of starting points and a proper choice of initial values, a global minimum region can be identified, often with multiple local minima. These approximate solutions within the same region have similar physical properties, indicating the robustness of the method. Occasionally, different regions may yield comparable $\chi^2$ values, but differ significantly in physical properties. In such cases, a physical argument is used to determine which quantities are more important to be more accurate. For example, two minimum regions may emerge, one with reasonable $T_{\chi}$ (according to the lattice) but low $m_\sigma$, and the other with larger physical $m_\sigma$ but high $T_{\chi}$. Prioritizing the importance of $T_{\chi}$ and the better description of the thermodynamics, the former parameter sets are preferred due to the broad resonance of $f_0(500)$.

In the present work, we extended the parameter set with $\xi_1$, $\xi_1^{1}$, and $\xi_2$, and performed several parameterizations including at least one of these couplings with $10^6$ random starting points in the parameter space. 
To show the qualitatively different scenarios, the average $\chi^2$ and $\chi^2_\text{red}$ values of the 100 best fits for each case except $\xi_X = \{\xi_1^1, \xi_2\}$ are shown in Tab.~\ref{tab:chisquares}.
\begin{table}[]
    \centering
    \begin{tabular}{l| c c c}
        \makecell{nonzero\\params.}~~ & $\bar{\chi}^2$ & $\bar{\chi}^2_{red}$ & ~~$\bar{\xi}_\text{eff}$~[GeV]\\
         \hline
         \hline
       $\xi_1$  & ~31.31~ & ~2.09~ & 1.507 \\
       $\xi_1^{1}$ & 29.70 & 1.98 & 1.537 \\ 
       $\xi_2$ & 33.50 & 2.23 & 1.589 \\ 
       $\xi_1$, $\xi_1^{1}$~ & 29.51 & 2.11 & 1.506 \\
       $\xi_1$, $\xi_2$ & 30.90 & 2.21 & 1.505 \\ 
       $\xi_1$, $\xi_1^{1}$, $\xi_2$ & 30.81 & 2.37  & 1.532 
    \end{tabular}
    \caption{The averaged $\chi^2$ and $\chi^2_{red}$ values of the best 100 fits with $10^6$ random starting points for different implementations of the $U(1)_A$ anomaly and the averaged effective anomaly coupling $\xi_\text{eff}$ for each case.}
    \label{tab:chisquares}
\end{table}
It was found that the lower dimensional anomaly terms give a slightly better $\chi^2_\text{red}$ value when only one of them is considered, but the difference is not large enough to exclude the higher dimensional contributions. Furthermore, we have found that when multiple terms are included, the parameterization usually favors solutions where there is only one dominant coupling and the rest are subleading, with only a few exceptional cases that show comparable $\chi^2_\text{red}$. It is also interesting to note that allowing some of the couplings for the anomaly terms to be negative would further reduce the $\chi^2$ of the fit, but also give rise to non-physical thermal behavior such as negative meson mass squares near $T_{\chi}$ or the absence of the phase transition. Therefore, not only for conceptual but also for physical reasons, we restrict the parameter space to $\xi_1,~\xi_1^1,~\xi_2>0$. 

The last column of Tab.~\ref{tab:chisquares} shows the average value of the effective anomaly coupling, which is defined as 
\begin{align}
\xi_\text{eff}\equiv\xi_1 + \xi_1^1 \left(\phi_N^2+\phi_S^2\right)/2+\xi_2 \phi_N^2 \phi_S / \sqrt{2} \text{ ,}
\end{align}
which results from the difference of $m_\pi^2$ and $m_{\eta_N}^2$ (explicitly shown in App. ~\ref{sec:elsm_details}), distinguished in the model only by the axial anomaly. Alternatively, one could use the form of the effective potential, which would differ only in the term with $\xi_2$ by a multiplicative factor of four. However, since the meson masses are used in the parameterization of the model, they better reflect the relationship between the different couplings. Finally, since the mixed term has a qualitatively similar analytical structure to the $Q=1$ term, we will concentrate on the cases with $\xi_1^1=0$ for simplicity.

\section{The Columbia plot} \label{sec:columbia}

\begin{figure*}[t]
    \centering
    \includegraphics[width=.32\textwidth]{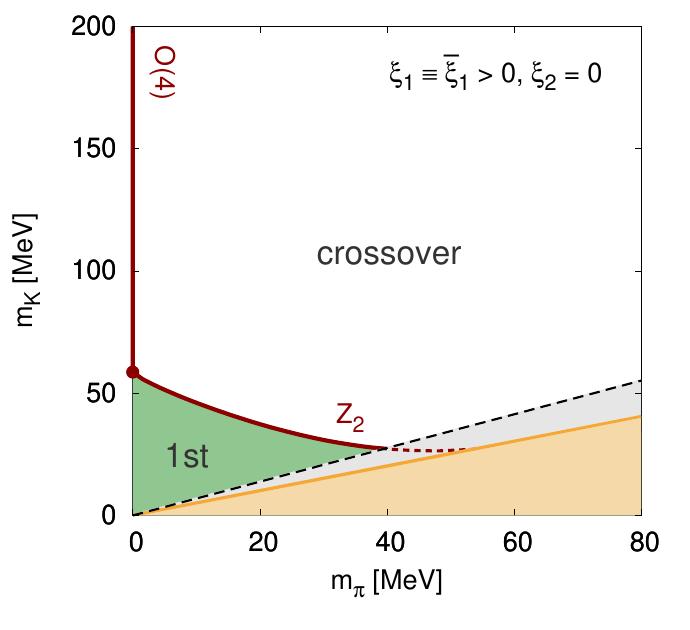}
    \includegraphics[width=.32\textwidth]{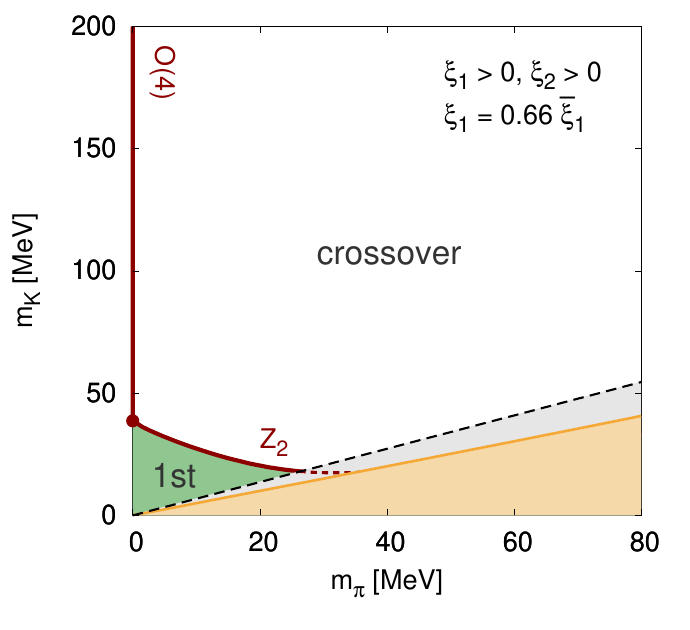}
    \includegraphics[width=.32\textwidth]{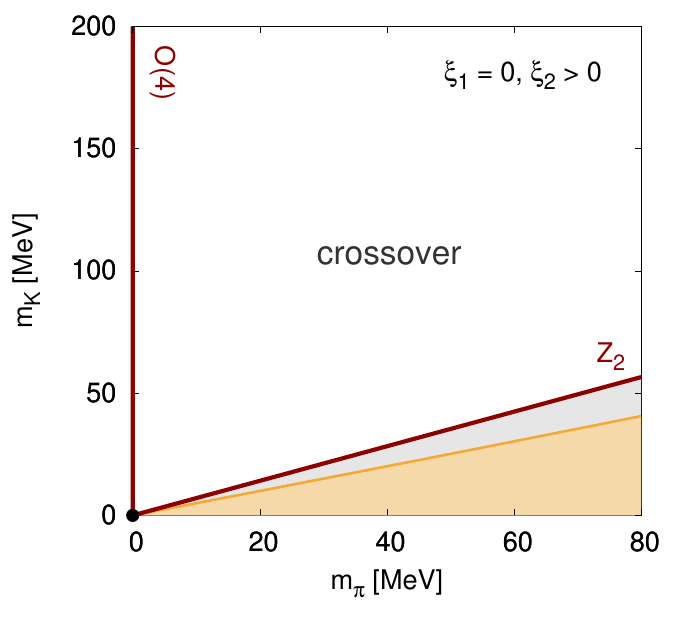}
    \caption{The Columbia plot shown in the $m_\pi$-$m_K$ plane for three independent parameterizations that has $\xi_1$ (left), $\xi_2$ (right), or both anomaly terms (center). The couplings of the other anomaly terms are set to zero and only $\phi_N$, $\phi_S>0$ are considered. The numerical values are $\xi_1 =\bar{\xi}_1=1.50$ GeV on the left, $\xi_2=278.79$ $\rm GeV^{-2}$ in the middle, and $\xi_2=1093.00$ $\rm GeV^{-2}$ on the right plot. The red lines represent the second order phase transition, the black dashed line shows $h_S=0$ and the gray area corresponds to $h_S<0$, where strange quarks have negative mass. Below the golden line $CP$ symmetry is spontaneously broken.}
    \label{Fig:Columbia_plot_xi1_xi2}
\end{figure*}

\begin{figure}[b]
    \centering
    \includegraphics[width=.45\textwidth]{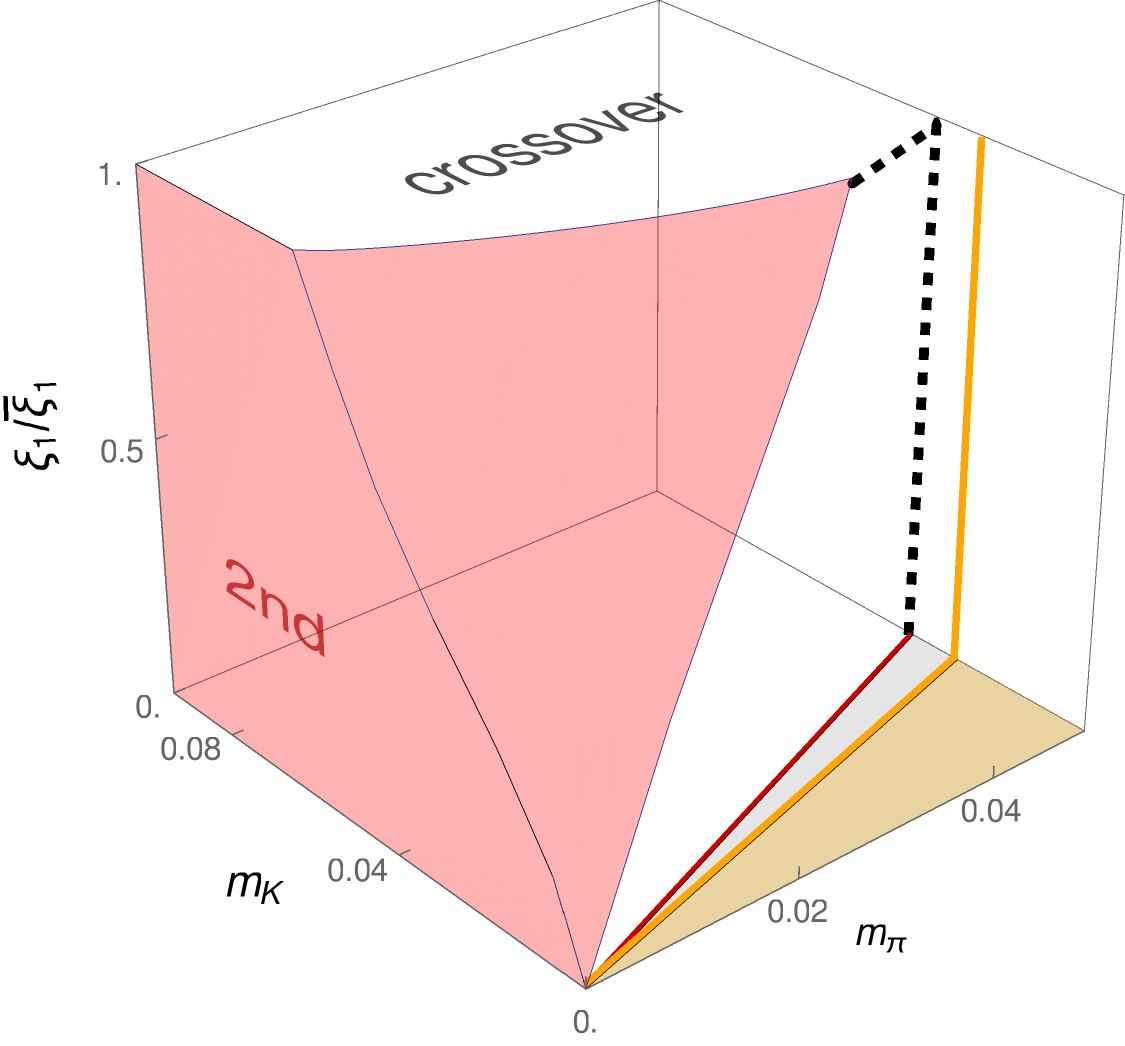}
    \caption{The 3D Columbia plot with the $\xi_1/\bar{\xi}_1$ relative weights shown along the z-axis  with fixed $\xi_\text{eff} =\bar{\xi}_1=1.50$ GeV effective coupling. $\xi_1^1=0$ and $\phi_N$, $\phi_S>0$. On the bottom $m_{\pi}\!-\!m_K$ plane $\xi_1=0$, while on the top $\xi_2=0$.}
    \label{Fig:3D_Columbia_plot_xi1_xi2}
\end{figure}

With viable sets of parameters regarding vacuum phenomenology at hand we can explore the resulting phase transition. To this end, we determine the light and strange chiral condensates from the solution of the equations of motion at finite temperature and read-off the order of phase transition for various values of the explicit symmetry breaking parameters $h_N$ and $h_S$, cf. \Eq{eq:HD}. Since these parameters can be related to current masses of light and strange quarks, $h_N\sim m_n$ and $h_S \sim m_s$, this gives us direct access to the Columbia plot.

We emphasize that we only vary $h_N$ and $h_S$ and keep all other parameters fixed. Since we are working with a low-energy model, this is not directly equivalent to varying the current quark masses in QCD. From a microscopic point of view, these masses should be fixed at some high-energy scale $\Lambda_{\rm UV} \gg 1\,{\rm GeV}$ in the perturbative regime of QCD. The parameters of our low-energy model could then be extracted by integrating-out fluctuations between $\Lambda_{\rm UV}$ and $\Lambda_{\rm eLSM}\lesssim 1\,{\rm GeV}$, which is the upper scale of validity of the model, see, e.g., Refs.\  \cite{Rennecke:2015lur, Springer:2016cji, Fu:2019hdw}. Different quark masses at $\Lambda_{\rm UV}$ would then not only lead to different $h_{N,S}$, but other model parameters would change as well at $\Lambda_{\rm eLSM}$. At the physical point we can, as in the present work, rely on experimentally measured properties of hadrons. Away from the physical point, the need for fundamental QCD input towards the chiral limit can, at least to some extent, be circumvented by using results from chiral perturbation theory, as in Refs.\ \cite{Herpay:2005yr, Kovacs:2006ym, Resch:2017vjs}. However, this concerns quantitative effects which are not the scope of the present work. 

The explicit symmetry breaking parameters directly determine the pion and kaon masses through the following expressions (the first is the realization of the Goldstone theorem in eLSM),
\begin{align}
\begin{split} \label{Eq:fpi_fK}
    f_\pi\, m_\pi^2 &=Z_{\pi} h_N\,,\\
    f_K\, m_K^2 &= Z_{K}\left(\frac{1}{2}h_N + \frac{1}{\sqrt{2}} h_S \right)\,,
    \end{split}
\end{align}
where $f_\pi=\phi_N/Z_\pi$ and $f_K=(\phi_N+\sqrt{2}\phi_S)/(2Z_K)$ are the pion and kaon decay constants. The wave function renormalizations $Z_{\pi,K}$ arise from meson mixing and are defined in Eqs.\ \eq{Eq:Z_pi} and \eq{Eq:Z_K}. We will express the Columbia plot in terms of these masses. Note that the kaon mass is always non-zero when $h_N > 0$, so non-zero pion masses always imply nonzero kaon masses. While we do not implement direct information from chiral perturbation theory away from the physical point, we only choose parameter sets that give reasonable values for the pion decay constant in the chiral limit $f_\pi \gtrsim 65$ MeV. This way, we use parameter sets that are at least roughly compatible with chiral perturbation theory. This is non trivial, since naive parameter fixing at the physical point can lead to a much too small and even vanishing $f_\pi$ in the three-flavor chiral limit \cite{Resch:2017vjs}.

In Fig.\ \ref{Fig:Columbia_plot_xi1_xi2} we show the resulting Columbia plots for three different, representative parameter sets. Note that we only show results with $\xi_1^{1} = 0$, since this coupling is only of quantitative, not qualitative relevance. The black dashed line corresponds to $h_S = 0$ which, as pointed out above, corresponds to $m_K > 0$ if $h_N >0$. The gray region is thus excluded for non-negative strange quark masses, i.e. for $h_S \geq 0$ (or, as explained below, for topological angle $\theta = 0$). 

The left figure shows the Columbia plot where the anomaly is implemented only with the conventional 't Hooft determinant, i.e.\ $\xi_1 \equiv \bar{\xi}_1 > 0$ and $\xi_2 = 0$. This is in qualitative agreement with other mean-field results where the anomaly is implemented in the same way, see, e.g., Ref.\ \cite{Resch:2017vjs}. In the light chiral limit, i.e.\ $m_\pi=0$, and with physical kaon masses, $m_K = m_K^{\rm phys}$, we find a second order transition. This extends to smaller values of $m_K$ until it finally ends at a tricritical point. Below this point we find an extended region where the chiral transition is of first order. From the tricritical point, a second order transition continues with decreasing $m_K$ and increasing $m_\pi$ until it ends at $h_S = 0$ (corresponding to a nonzero $m_K$). For $h_S = 0$ and even larger $m_\pi$, only a crossover is found. This is consistent with the expectation for the one-flavor limit, which is reached for $h_N \rightarrow \infty$ and $h_S$ finite. In this case, $\xi_1$ explicitly breaks the chiral symmetry and the transition is a crossover.

In the middle plot of Fig.~\ref{Fig:Columbia_plot_xi1_xi2} we show the result for a reduced value of the 't Hooft coupling compared to the left plot, $\xi_1 = 0.66\, \bar{\xi}_1$. To be able to describe the vacuum phenomenology at the physical point, especially regarding anomalous quantities like the $\eta^\prime$ mass, this forces the higher-order anomalous coupling to be nonzero, $\xi_2 > 0$. As shown in \Tab{tab:chisquares}, this scenario is still phenomenologically viable. The resulting Columbia plot is qualitatively the same as for $\xi_2 = 0$, but the size of the first-order region in the lower-left corner has shrunk considerably.

The right plot of Fig.\ \ref{Fig:Columbia_plot_xi1_xi2} shows the extreme scenario where $\xi_1 = 0$, so that all anomalous effects are entirely encoded in $\xi_2 > 0$. In this case the first-order region is completely gone, and there is a second-order phase transition for $m_\pi = 0$ and all $m_K \geq 0$. Interestingly, the crossover along the line $h_S = 0$ becomes a second-order transition. This is not entirely unexpected, since $\xi_1$ gives rise to a non-zero chiral condensate in the one-flavor limit. The choice $\xi_1 = 0$ thus facilitates a phase transition rather than a crossover in this limit, although $U(1)_A$ remains broken. Note that according to the above, the second order phase transition in the one-flavor chiral limit\footnote{The $h_S = 0$, $h_N\to\infty$ (i.e. $m_\pi \to \infty$) limit on the Columbia plot, which corresponds to the lower right corner when shown in $h_N$ and $h_S$ instead of the pion and kaon masses.} is continuously connected to the second order transition in the three-flavor chiral limit (lower left corner of the Columbia plot) in this scenario. 

For a better visualization of the above scenarios with $\xi_1$ and $\xi_2$, one can obtain a chain of parameter sets that interpolates between the $\xi_1=0$ and $\xi_2=0$ cases, keeping the $\xi_\text{eff}$ fixed and fitting the same physical quantities in each step. Such a set of parameterizations is of course not unique, but can be defined systematically. We created a chain with $\xi_\text{eff}=1.50$ GeV (starting from the parameter set of the leftmost figure in Fig.\eqref{Fig:Columbia_plot_xi1_xi2}) and the resulting 3-dimensional Columbia plot is shown in Fig.~\ref{Fig:3D_Columbia_plot_xi1_xi2}. It can be seen that the size of the first-order region decreases with decreasing $\xi_1$ from its maximum value until it vanishes. This change is generally monotonic.

It is interesting to note that the second-order line with $Z_2$ universality class -- which separates the first-order and crossover regions for finite $h_N$ and $\xi_1>0$ -- goes into the $h_S<0$ region (thus below the dashed $h_S=0$ line). If only the terms with $\xi_1$ and $\xi_2$ couplings are used and the relative weight of $\xi_1$ decreases, this line transforms into the line of $h_S=0$. Therefore the diagonal $Z_2$ segment in the rightmost plot of Fig.~\ref{Fig:Columbia_plot_xi1_xi2} (and the bottom of Fig.~\ref{Fig:3D_Columbia_plot_xi1_xi2}) is not separated from the $Z_2$ line of the $\xi_1>0$ cases, but is part of the same second order surface in the 3D plot (shown in red), part of which is not shown because it is in the $h_S<0$ region.

We emphasize that, as discussed in the introduction, there is currently no evidence in the critical physics literature for a second-order transition in the three-flavor chiral limit if $U(1)_A$ remains broken. In the present case, the second-order transition arises because the lowest order anomalous interaction $\xi_1$ vanishes, not because $U(1)_A$ is restored. In fact, since a second-order transition is also found in the mean-field approximation when all anomalous correlations vanish \cite{Resch:2017vjs}, we conclude that the order of the transition in the three-flavor chiral limit is completely controlled by the 't Hooft determinant: the smaller $\xi_1$, the smaller the first-order region; a second-order transition in the three-flavor chiral limit is only possible when $\xi_1$ vanishes identically.

As pointed out in Ref.\ \cite{Pisarski:2024esv}, this latter result follows from the general expectation that a non-zero cubic term \mbox{$\xi_i\, (\det\Phi + \det\Phi^\dagger)$} will always spoil the second order transition of $\phi^4$ theory, since it dominates the infrared behavior near the phase transition. Here we find more generally that the size of the lowest order anomalous coupling controls the size of the first order region.

\section{Spontaneous $CP$ violation}

In the regime of negative strange quark mass, the gray region in Figs.~\ref{Fig:Columbia_plot_xi1_xi2} and \ref{Fig:3D_Columbia_plot_xi1_xi2}, we find a transition to a phase with non-vanishing $\eta^\prime$ condensate, where parity ($P$) is hence spontaneously broken. Since the grand potential is still invariant under charge conjugation ($C$) this implies spontaneous $CP$ breaking as well. The transition line, where $m_{\eta^\prime}=0$, is marked by the golden line in Figs.~\ref{Fig:Columbia_plot_xi1_xi2} and \ref{Fig:3D_Columbia_plot_xi1_xi2}. The $CP$-broken phase occurs below this line in the gold area. 
Since the model parameters, and hence the effective anomaly coupling, are fitted by the meson masses, this line almost coincides for the different implementations of the anomaly if the given parameterizations are close to each other in parameter space.

Note that we define $\eta^\prime$ for general $h_N$ and $h_S$ as the predominantly strange physical pseudoscalar isoscalar state, which is the combination with the minus sign in Eq.~\eqref{Eq:NS_to_physical_mass} for $h_S<h_N$ (see App.~\ref{sec:elsm_details} for details). Furthermore, we consider only positive $\phi_N$ and $\phi_S$, but remark that for the case where $\xi_2$ is the only nonzero anomalous coupling, the solution for $\phi_S > 0$ is a local minimum. The global minimum is found at $\phi_S < 0$, which we exclude here for physical reasons\footnote{Unlike the positive solution at the local minimum for $h_S < 0$, the $\phi_S<0$ solution is not continuously connected to the global solution for $h_S > 0$. Hence, if $\xi_2$ is the only anomalous term there is a first-order phase transition at $h_S = 0$. This is expected, as in this case the $-h_S \phi_S$ term of the effective potential (see Eq.~\eqref{Eq:GrandPotCl}) is the only odd term in $\phi_S$. 
This solution gives a relatively high kaon mass, outside the range of the Figs.~\ref{Fig:Columbia_plot_xi1_xi2} and \ref{Fig:3D_Columbia_plot_xi1_xi2}. We therefore restrict ourselves to $\phi_S \geq 0$, but note that that $\phi_S < 0$ solution only leads to quantitative changes. The qualitative features of our resuts, in particular the occurence of spontaneous $CP$ breaking, remain the same.}. 

The spontaneous breaking of $CP$ found here can be viewed as a realization of Dashen's phenomenon \cite{Dashen:1970et, Witten:1980sp}. To see this, we introduce the topological $\theta$ angle into our model. Following Ref.\ \cite{tHooft:1986ooh}, this can be done by noting that $\theta$ enters as the coefficient of the topological charge density in the QCD action. Hence, each operator from a sector with topological charge $Q$ will come with an additional factor $e^{i Q \theta}$ \cite{Rennecke:2020zgb}, so that only the anomalous interactions in \Eq{eq:Lqu} are modified,
\begin{align} \label{eq:Lquthet}
    \begin{split}
        \mathcal{L}_{\rm qu}(\theta) &= -\xi_1 \Big(e^{i\theta}\det\Phi + e^{-i\theta}\det\Phi^\dagger\Big)\\ 
        &\quad - \xi_1^{1}\, {\rm tr}\big( \Phi^\dagger\Phi\big) \Big(e^{i\theta}\det\Phi + e^{-i\theta}\det\Phi^\dagger\Big)\\
        &\quad-\xi_2 \Big[e^{i2\theta}\big(\det\Phi\big)^2 + e^{-i2\theta}\big(\det\Phi^\dagger\big)^2\Big]\,.
    \end{split}
\end{align}
Performing an axial rotation with angle $\alpha$ on a single quark flavor, e.g., the strange quark,
\begin{align}\label{eq:sax}
    s \rightarrow e^{i \frac{\alpha}{2} \gamma_5} s\,, 
\end{align}
we find
\begin{align}
    \mathcal{L}_{\rm qu}(\theta) \rightarrow \mathcal{L}_{\rm qu}(\theta-\alpha)\,. 
\end{align}
It is well known that by an appropriate choice of the axial rotation, in this case $\alpha = \theta$, the $\theta$-dependence can be ``moved" from the anomalous interactions to the quark mass. In particular, under the transformation \eq{eq:sax} the explicit symmetry breaking term in strange direction then is for $\alpha = \theta$,
\begin{align}
- h_S \phi_S \rightarrow - h_S \big(\phi_S \cos\theta + \eta_S \sin\theta \big)\,,
\end{align}
where $\phi_S \sim \bar s s$ and $\eta_S \sim i \bar s\gamma_5 s$. Thus, by setting $\theta = \pi$ we can flip the sign of the symmetry breaking term. The gray region in Figs.\ \ref{Fig:Columbia_plot_xi1_xi2} and \ref{Fig:3D_Columbia_plot_xi1_xi2} can therefore be viewed as part of a phase diagram where $\theta = \pi$.

Recall that $\theta \rightarrow -\theta$ under $CP$ transformations, so the system is $CP$ symmetric for $\theta = 0, \pi$. While at $\theta =\pi$ no symmetries are broken explicitly, Dashen observed that $CP$ can be broken spontaneously in this case \cite{Dashen:1970et, Witten:1980sp}. This is exactly what we see here as well. We note that the case where the 't Hooft term is the only anomalous interaction has already been investigated in detail in Nambu--Jona-Lasinio models in mean-field approximation, see, e.g., Refs.\ \cite{Fujihara:2005wk, Boer:2008ct, Sakai:2011gs}. On the mean-field level, we do not expect any relevant qualitative differences between these models and ours for this discussion. 
However, higher order anomalous couplings have not been considered before in this context. It is noteworthy that while $\theta = \pi$ effectively changes the sign of $\xi_1$ and $\xi_1^1$, it does not affect $\xi_2$. We find an $\eta^\prime$ condensate in any case.

\section{Conclusions}

The nature of the chiral phase transition, especially in the three-flavor chiral limit, is not yet fully understood. And a mean-field analysis of a low-energy effective model certainly cannot solve this open problem. But it can tell us about important qualitative features. We have found that different realizations of the axial anomaly in terms of anomalous meson correlations are phenomenologically viable: whether we break $U(1)_A$ with the conventional 't~Hooft determinant, anomalous higher order operators, or a combination of them, the resulting eLSM accurately describes a large number of physical quantities in vacuum.

However, we found that the exact realization of the axial anomaly critically affects the chiral phase transition. The strength of the 't Hooft determinant at $T_\chi$
\footnote{In our case the coefficients of the anomaly terms are temperature independent, so their relative strength is the same at $T=T_\chi$ as at $T=0$, where the parameterization is performed.}
directly controls the size of the first order region in the lower left corner of the Columbia plot. In our analysis, a second-order transition in the three-flavor limit required a vanishing 't~Hooft determinant at $T_\chi$. We emphasize that this does not necessarily imply the restoration of $U(1)_A$, as it can still be encoded in higher order operators. This is in line with Ref.~\cite{Pisarski:2024esv}, where it was conjectured that a small or even vanishing first-order region arises from the \emph{effective} coupling $\xi_\text{eff}$ becoming small or even zero at $T_\chi$. It has been argued that anomalous couplings generated by multi-instantons can contribute to $\xi_\text{eff}$ only with appropriate (positive) powers of the chiral condensate, since they are of higher order. Thus, the single-instanton contribution, which corresponds to the 't Hooft determinant in the semiclassical (large-$T$) limit, must be suppressed against these higher-order contributions if $\xi_1$ is small at $T_\chi$, where the chiral condensate vanishes.

Of course, the exact strength and temperature dependence of the anomalous couplings cannot be deduced from our analysis. It would be worthwhile to extract these couplings from microscopic computations on the lattice or with functional methods. Furthermore, in the mean field we can find a second order transition in the chiral transition, independent of the fate of $U(1)_A$ at $T_\chi$. As discussed in the introduction, the situation in the realistic case beyond the mean-field is not yet fully resolved. So far, a second order transition seemed to be possible only if $U(1)_A$ restores at $T_\chi$, which implies that all anomalous couplings must vanish, not only $\xi_1$. However, the case where $U(1)_A$ remains broken at $T_\chi$ has not been systematically studied. If it turns out that there is a stable fixed point in this case, our analysis suggests a vanishing critical value of $\xi_1$ (which has mass dimension one), and nonzero critical values for at least some of the higher order couplings $\xi_{n>1}$.

\acknowledgments

We acknowledge discussions with Christian Fischer, Owe Philipsen, Bernd-Jochen Schaefer, Lorenz von Smekal, and Gy\"orgy Wolf. 
R.D.P. is supported by the U.S. Department of Energy under contract DE-SC0012704, and thanks the Alexander v. Humboldt Foundation for their support. F.~R.\ is grateful for the hospitality of the Wigner Research Centre for Physics, where this work has been initiated.
F.\,R.\ is supported by the Deutsche Forschungsgemeinschaft (DFG, German Research Foundation) through the Collaborative Research Center TransRegio CRC-TR 211 “Strong- interaction matter under extreme conditions” – project number 315477589 – TRR 211. P.~K. and G.~K. acknowledge support by the Hungarian National Research, Development and Innovation Fund under Project number K 138277. The work of G.~K. is partially supported by the Polish National Science Centre (NCN) under OPUS Grant
No. 2022/45/B/ST2/01527.

\appendix

\section{Details of the eLSM} \label{sec:elsm_details}

As we discussed in Section \ref{sec:LSM}, the eLSM Lagrangian can be written as $\mathcal{L}_{\rm cl}+ \mathcal{L}_{\rm esb}+ \mathcal{L}_{\rm qu}$. According to Refs.~\cite{Kovacs:2016juc,Kovacs:2021kas,Giacosa:2024epf}, the chirally symmetric part reads 
\begin{widetext}
\begin{align}
    \mathcal{L}_{\rm cl} =& \Tr \left[ \left( D_\mu \phi \right)^\dagger \left( D^\mu \phi \right) \right] - m_0 \Tr \left( \phi^\dagger \phi \right) - \lambda_1 \left[ Tr \left( \phi^\dagger \phi \right) \right]^2  - \lambda_2 \left[ \Tr \left( \phi^\dagger \phi \right)^2 \right] \nonumber  - \frac{1}{4} \Tr \left[ L_{\mu\nu}L^{\mu\nu}+R_{\mu\nu}R^{\mu\nu} \right] \nonumber \\
    &+ 2 h_3 \Tr \left[ R_\mu \phi^\dagger L^\mu \phi \right]  +\frac{h_1}{2} \Tr \left( \phi^\dagger \phi \right) \Tr \left[ L_\mu L^\mu +R_\mu R^\mu \right] + h_2 \Tr \left[ \left( \phi R_\mu \right)^\dagger \left( \phi R^\mu \right) + \left( L_\mu  \phi \right)^\dagger \left( L^\mu \phi  \right) \right] \label{eq:Lcl}\\ 
    & + \Tr \left[\left(\frac{m_1^2 }{2} \right) \left( L_\mu L^\mu + R_\mu R^\mu \right) \right] - \frac{g_2}{2} \left( \Tr \lbrace L_{\mu\nu} \left[ L^\mu , L^\nu \right] \rbrace + \Tr \lbrace R_{\mu\nu} \left[ R^\mu , R^\nu \right] \rbrace \right) \nonumber\\
    &+\bar\psi \left[ i \gamma_\mu \partial^\mu -g_S \left( S-i\gamma_5P\right)-g_V\gamma^\mu \left(V_\mu + \gamma_5 A_\mu \right)\right] \psi , \nonumber
\end{align}
\end{widetext}
where the scalar, pseudoscalar, and the left- and right-handed vector nonets are
\begin{align}
    \phi =& S+i P=\sum_a \left(S_a + i P_a\right)T_a,  \nonumber\\
    L^\mu =&\sum_a \left(V^\mu_a +A^\mu_a \right)T_a, \;
	R^\mu =\sum_a \left(V^\mu_a -A^\mu_a \right)T_a. \label{Eq:mfields}
\end{align}
Here $T_a=\lambda_a/2$ are the generators of the $U(3)$ group defined by the $\lambda_i~(i=1,\ldots,8)$ Gell-Mann matrices and $\lambda_0=\sqrt{2/3}~\mathbb{1}_{3\times3}$.
In the fermionic part $\psi=(q_u,q_d,q_s)^T$ are the constituent quarks, and for the present case the (axial) vector-fermion interaction is not considered, hence $g_V=0$. The covariant derivative and the field strength tensors appearing in \eqref{eq:Lcl} can be written as
\begin{align}
    &D^\mu \phi= \partial^\mu \phi - i  g_1(L^\mu \phi -\phi R^\mu), \nonumber\\
    L^{\mu\nu} =& \partial^\mu L^\nu -\partial^\nu L^\mu, \quad
    R^{\mu\nu} = \partial^\mu R^\nu -\partial^\nu R^\mu ,
\end{align} 
where we have omitted the currently irrelevant electromagnetic field.
As discussed in \cite{Kovacs:2021kas}, the covariant derivative could be written in a simplified form, but it can be shown -- with redefinition of certain coefficients of the Lagrangian -- to be equivalent to the form presented above. Explicit symmetry breaking is implemented by the contributions
\begin{align} \label{eq:Lesb}
    \mathcal{L}_{\rm esb} = \Tr \left[ H \left( \phi + \phi^\dagger \right) \right] + \Tr \left[\Delta \left( L_\mu L^\mu + R_\mu R^\mu \right) \right] \text{ ,}
\end{align}
where the external fields are defined as
\begin{align}\label{eq:HD}
H&=H_0 T_0 + H_8 T_8 = \frac{1}{2} \rm{diag} \left( h_{N},h_{N},\sqrt{2}h_{S} \right),\\
\Delta&=\Delta_0 T_0 + \Delta_8 T_8 = \rm{diag} \left( \delta_{N},\delta_{N},\delta_{S} \right) .
\end{align}
Note that the terms with $\delta_{N},\delta_{S}$ are not present in the field equations (either explicitly or implicitly), so they do not affect the phase transition. The anomaly breaking $\mathcal{L}_{\rm qu}$ contribution is shown in Eq.~\eqref{eq:Lqu}. The parameters of the model to be determined with the parameterization are $\phi_{N}$, $\phi_{S}$ (equivalent to using $h_N$ and $h_S$ which are now determined via the field equations), $m_{0}^2$, $m_{1}^2$, $\lambda_{1}$, $\lambda_{2}$, $\delta_{S}$, $g_{1}$, $g_{2}$, $h_{1}$, $h_{2}$, $h_{3}$, $g_{S}$, $M_0$, and those from $\mathcal{L}_{\rm qu}$. Note that there are other terms allowed by the symmetry constraints that contain only vector and axial vector fields, but these would not contribute to any of the physical quantities considered. The parameters ($m_{1}^2$, $\delta_{S}$, $g_{1}$, $g_{2}$, $h_{1}$, $h_{2}$ and $h_{3}$) coupled to terms containing (axial) vector fields appear in the masses of these mesons and in some of the decay widths. However, via the (pseudo)scalar-(axial) vector mixing, they also directly affect the (pseudo)scalar meson masses and thus the Columbia plot when depicted in $m_\pi$ and $m_K$.

The grand canonical potential is calculated at the mean-field level and reads
\begin{align} \label{Eq:GrandPot}
    \Omega = V_\text{cl} + \Omega_{\bar{q}q}^\text{v} + \Omega_{\bar{q}q}^T\ ,
\end{align}
where the first term is the tree-level, classical contribution
\begin{align} \label{Eq:GrandPotCl}
    V_\text{cl} =& \frac{m_0^2}{2} \left(\phi_N^2+\phi_S^2\right) - h_N \phi_N - h_S \phi_S \nonumber\\
    &+\frac{\lambda_1}{4} \left(\phi_N^2 + \phi_S^2\right)^2 + \frac{\lambda_2}{8} \left( \phi_N^4 + 2\phi_S^4\right) \\
    &-\frac{\phi_N^2 \phi_S}{2\sqrt{2}} \left( \xi_1 + \frac{\xi_1^1}{2} \left( \phi_N^2 + \phi_S^2 \right) \right) - \frac{\xi_2}{16} \phi_N^4 \phi_S^2. \nonumber
\end{align}
The second and third terms in Eq.~\eqref{Eq:GrandPot} are the renormalized fermionic vacuum and the fermionic thermal fluctuations
\begin{align}
    \Omega_{\bar{q}q}^\text{v} =& -\frac{N_c}{(4\pi)^2} \sum_{f} m_f^4 \ln \frac{m_f^2}{M_0^2}, \\
    \Omega_{\bar{q}q}^\text{T} =& -2T \sum_{f} \int \frac{d^3p}{(2\pi)^3} \left[\ln g_f^+ (p) + \ln g_f^- (p) \right], 
\end{align}
where $M_0$ is the renormalization scale. In the absence of the Polyakov loop $g_f^\pm = 1 - e^{-N_c\beta (E_f(p)\mp\mu_q)}$ with $E_f^2(p)=p^2+m_f^2$ (the Polyakov loop modified version of $g_f^\pm$ at general $N_c$ can be found in \cite{Kovacs:2023kbv}), where $m_f$ is the constituent quark mass being $m_{u,d}=g_S \phi_N / 2$ for the light quark flavors and $m_{s}=g_S \phi_S / \sqrt{2}$ for the strange quark.

The field equations can be obtained by minimizing the grand potential in the order parameters. For $\phi_N$ and $\phi_S$ one finds
\begin{align} 
    h_N&=m_0^2 \phi_N + \Big(\lambda_1 + \frac{\lambda_2}{2} \Big)\phi_N^3 + \lambda_1\phi_N\phi_S^2 -\frac{\xi_2}{4} \phi_N^3 \phi_S^2 \nonumber\\& -
    \frac{\phi_N \phi_S}{\sqrt{2}}\Big(\xi_1 + \xi_1^1\big(\phi_N^2 + \frac{\phi_S^2}{2}\big)\Big) + \frac{g_S}{2} \sum_{l=u,d} \langle \bar{q_l} q_l\rangle , \label{Eq:FE_phiN}\\
    h_S&= m_0^2 \phi_S + \left(\lambda_1 + \lambda_2 \right)\phi_S^3 + \lambda_1\phi_N^2\phi_S - \frac{\xi_2}{8} \phi_N^4\phi_S \nonumber \\ &-\frac{\phi_N^2}{2\sqrt{2}}\Big(\xi_1 + \frac{\xi_1^1}{2}\big( \phi_N^2 + 3\phi_S^2 \big)\Big) + \frac{g_S}{\sqrt{2}} \langle \bar{q_s} q_s\rangle . \label{Eq:FE_phiS}
\end{align}
Here we have introduced the quark-antiquark condensates $\langle\bar{q}_f q_f\rangle=-4N_c m_f \mathcal{T}_f$, which can be defined with the tadpole integral
\begin{align}
    \mathcal{T}_f =& \frac{m_f^2}{32\pi^2} \Big(1+2\ln \frac{m_f^2}{M_0^2}\Big) \nonumber \\
    &- \int \frac{d^3p}{(2\pi)^3} \frac{1}{2E_f(p)} \left(f_f^+(p) + f_f^-(p)\right),
\end{align}
containing the $f_f^\pm$ (possibly Polyakov loop modified) Fermi-Dirac distributions for particles and antiparticles.

The curvature meson masses are calculated as the second derivative of the grand potential with respect to the fluctuating mesonic fields. In the tree-level contribution, however, there is a mixing between the scalar and the vector mesons for each field, and between the pseudoscalar and axial-vector mesons for the isospin doublets (i.e., for $K_0^\star$ and $K_1^{\star\, \mu}$) \footnote{It can be shown, that the nonpropagating 4-longitudinal modes of the (axial) vectors participate in the mixing \cite{Kovacs:2021kas}.}. When the masses of the physical states are resolved, a wavefunction renormalization factor for the (pseudo)scalars naturally emerges \cite{Kovacs:2021kas}, which reads
\begin{align}
    Z_\pi^2 = &Z_{\eta_N}^2=\frac{m_{a_1}^2 }{m_{a_1}^2-g_1^2 \phi_N^2},\label{Eq:Z_pi}\\
    Z_K^2 =& \frac{4 m_{K_1}^2 }{ 4 m_{K_1}^2-g_1^2 (\phi_N+\sqrt{2}\phi_S)^2}, \label{Eq:Z_K}\\
    Z_{\eta_S}^2 =& \frac{m_{f_{1S}}^2 }{m_{f_{1S}}^2-2g_1^2 \phi_S^2},\\
    Z_{K_0^\star}^2 =& \frac{4 m_{K^\star}^2 }{ 4 m_{K^\star}^2-g_1^2 (\phi_N-\sqrt{2}\phi_S)^2},
\end{align}
where the $m_{a_1}$, $m_{K_1}$, $m_{f_1^S}$, and $m_{K^\star}$ are the masses of the corresponding (axial) vector mesons and their explicit form can be found in \cite{Parganlija:2012fy}.\footnote{If the (axial) vector-fermion interactions are also taken into account the appearing mass corresponds to the 4-longitudinal mode \cite{Kovacs:2021kas}.}
With these, the masses of the (pseudo)scalars in the $N-S$ bases are given by
\begin{widetext}
\begin{align}
    m_\pi^2 =& Z_\pi^2\bigg[m_0^2+\Big(\lambda_1+\frac{\lambda_2}{2}\Big)\phi_N^2+\lambda_1\phi_S^2-\left(\xi_1 + \xi_1^1 \left(\phi_N^2 + \phi_S^2/2\right)\right)\frac{\phi_S}{\sqrt{2}} - \frac{\xi_2}{4} \phi_N^2\phi_S^2-2 N_c g_S^2~ \mathcal{T}_u \bigg] = Z_\pi^2 \frac{h_N}{\phi_N} , \label{Eq:m_pi}\\
    m_K^2 =& Z_K^2\bigg[m_0^2+\Big(\lambda_1+\frac{\lambda_2}{2}\Big)\phi^2_N+\left(\lambda_1+\lambda_2\right)\phi^2_S - \frac{\lambda_2}{\sqrt{2}}\phi_S\phi_N - \Big(\xi_1+ \frac{\xi_1^1}{2} \big(\phi_N^2 + \sqrt{2} \phi_N \phi_S + \phi_S^2 \big)\Big)\frac{\phi_N}{2} \nonumber \label{Eq:m_K}\\ 
    & \quad - \frac{\xi_2}{4\sqrt{2}}\phi_N^3 \phi_S -2 N_c g_S^2~ \frac{\phi_N \mathcal{T}_u+\sqrt{2}\phi_S \mathcal{T}_s}{\phi_N + \sqrt{2}\phi_S}\bigg] =Z_K^2 \frac{h_N+\sqrt{2}h_S}{\phi_N+\sqrt{2}\phi_S} ,\\
    m_{\eta_N}^2 = & Z_{\eta_N}^2\bigg[m_0^2+\Big(\lambda_1+\frac{\lambda_2}{2}\Big)\phi_N^2+\lambda_1\phi_S^2+\Big(\xi_1+\frac{\xi_1^1}{2} \phi_S^2\Big)\frac{\phi_S}{\sqrt{2}} + \frac{3}{4} \xi_2 \phi_N^2 \phi_S^2 - 2 N_c g_S^2~ \mathcal{T}_u\bigg] = \nonumber \\
    =& Z_{\eta_N}^2 \left[ \frac{h_N}{\phi_N} + \sqrt{2}\xi_1 \phi_S +\frac{\xi_1^1}{\sqrt{2}} \big(\phi_N^2 + \phi_S^2\big) \phi_S + \xi_2 \phi_N^2 \phi_S^2\right] \\
    m_{\eta_S}^2 = & Z_{\eta_S}^2\left[m_0^2+\lambda_1\phi_N^2+\left(\lambda_1+\lambda_2\right)\phi_S^2 - \frac{\xi_1^1}{2\sqrt{2}}\phi_N^2 \phi_S + \frac{\xi_2}{8} \phi_N^4  - 2 N_c g_S^2 \mathcal{T}_s \right] =\nonumber \\
    =& Z_{\eta_S}^2 \left[\frac{h_S}{\phi_S} + \sqrt{2}\left(\frac{\xi_1}{4}+\frac{\xi_1^1}{8} \big(\phi_N^2+ \phi_S^2\big)\right) \frac{\phi_N^2}{\phi_S} + \frac{\xi_2}{4} \phi_N^4 \right] \label{Eq:m_etaS}\\
    m_{\eta_{NS}}^2= & Z_{\eta_N} Z_{\eta_S}\left[\left(\xi_1 + \frac{\xi_1^1}{2} (\phi_N^2+\phi_S^2)\right)\frac{\phi_N}{\sqrt{2}} + \frac{\xi_2}{2} \phi_N^3 \phi_S \right] 
\end{align}
\begin{align}
    m_{a_0}^2=&m_0^2 + \left(\lambda_1+\frac{3\lambda_2}{2}\right) \phi_N^2 + \lambda_1 \phi_S^2 +\left(\xi_1 + \frac{\xi_1^1}{2} \phi_S^2\right)\frac{ \phi_S }{\sqrt{2}} + \frac{\xi_2}{4}\phi_N^2\phi_S^2 - 2 N_c g_S^2~\bigg(\mathcal{T}_u - \frac{m_u^2}{2} \mathcal{B}_u \bigg) \\
    m_{K_0^\star}^2=&Z_{K_0^\star}^2 \bigg[m_0^2 + \left(\lambda_1+\frac{\lambda_2}{2}\right) \phi_N^2  +\left(\lambda_1+\lambda_2\right) \phi_S^2 + \frac{\lambda_2}{\sqrt{2}} \phi_N \phi_S + \left(\xi_1 + \frac{\xi_1^1}{4}\big(\phi_N^2 -\sqrt{2} \phi_N\phi_S + \phi_S^2 \big)\right) \frac{\phi_N}{2} \nonumber \\ 
    & \quad + \frac{\xi_2}{4\sqrt{2}} \phi_N^3\phi_S -2 N_c g_S^2~ \frac{\phi_N \mathcal{T}_u-\sqrt{2}\phi_S \mathcal{T}_s}{\phi_N - \sqrt{2}\phi_S} \bigg]\\
    m_{\sigma_N}^2 =& m_0^2 + 3 \left(\lambda_1+\frac{\lambda_2}{2}\right)\phi_N^2 +\lambda_1 \phi_S^2 - \left( \xi_1 + \frac{\xi_1^1}{2}\big(6\phi_N^2 + \phi_S^2 \big) \right)\frac{\phi_S}{\sqrt{2}} - \frac{3}{4} \xi_2 \phi_N^2 \phi_S^2 - 2 N_c g_S^2~\bigg(\mathcal{T}_u - \frac{m_u^2}{2} \mathcal{B}_u \bigg)\label{Eq:m_sigN}\\
    m_{\sigma_S}^2=& m_0^2 + \lambda_1\phi_N^2 + 3 \left(\lambda_1+\lambda_2\right) \phi_S^2 -\frac{3}{2} \xi_1^1 \phi_N^2\phi_S - \frac{\xi_2}{8} \phi_N^4 - 2 N_c g_S^2~\big(\mathcal{T}_s - m_s^2 \mathcal{B}_s \big) \label{Eq:m_sigS}\\
    m_{\sigma_{NS}}^2 =& 2\lambda_1 \phi_N\phi_S - \left(\xi_1 + \xi_1^1 \left(\lambda_1+\frac{3\lambda_2}{2}\right) \right)\frac{\phi_N}{\sqrt{2}} - \frac{\xi_2}{2} \phi_N^3\phi_S \label{Eq:m_sigNS}
\end{align}
\end{widetext}
where $\mathcal{B}_f$ is the bubble integral defined as $\mathcal{B}_f=-\partial \mathcal{T}_f/m_f^2$. To diagonalize the isoscalar sector, one can use
\begin{align}
m_{\varphi_H,\varphi_L}^2=&\frac{1}{2} \left( m_{\varphi_N}^2 + m_{\varphi_S}^2 \pm \sqrt{\left( m_{\varphi_N}^2 - m_{\varphi_S}^2\right)^2 + 4m_{\varphi_{NS}}^4} \right)  \label{Eq:NS_to_physical_mass}
\end{align} 
for both the scalars ($\varphi=\sigma$) and the pseudoscalars ($\varphi=\eta$). 

Note that the assignment of physical fields for general $h_N$ and $h_S$ may depend on the choice of definition. If the distinction is based on the sign in Eq.~\eqref{Eq:NS_to_physical_mass}, and the combinations with larger masses are associated with $\eta^\prime$ and $f_0^H$, then at low $h_S$ one finds predominantly strange $\eta$ and $f_0^L$ mesons behaving as chiral partners. However, one could choose the strange-nonstrange content of the diagonalized fields (which could be quantified by mixing angles) and the behavior during chiral restoration to identify the physical states. In this case for $m_{\varphi_N}\lesssim m_{\varphi_S}$ one has $\sigma_H=f_0^H$, $\sigma_L=f_0^L$, $\eta_H=\eta'$, $\eta_L=\eta$, while for $m_{\varphi_N} \gtrsim m_{\varphi_S}$ (which is in the lower part of the Columbia plot) the opposite assignment holds (the cases are clearly separated if the mixing term is zero, otherwise the mixing angle can be used as a guide). To favor the more intuitive behavior near the transition, we use the latter definition in this paper.

As shown in Eqs.~\eqref{Eq:m_pi}-\eqref{Eq:m_etaS}, the pseudoscalar masses can be expressed in terms of $h_N$, $h_S$, and the couplings of the anomaly terms using Eqs.~\eqref{Eq:FE_phiN}-\eqref{Eq:FE_phiS}. 
These can be used to study the behavior of the meson masses in the different chiral limits.
At the left edge of the Columbia plot, where $h_N=0$, the pion mass vanishes at $T<T_\chi$ as expected. 
The mass of the pion and its chiral partner (the predominantly nonstrange physical state in the scalar isoscalar sector) become degenerated at $T=T_\chi$ in such a way that both masses vanish. Along with this, the order parameter goes to zero for $T\to T_\chi$ in a second order (or with $\xi_1>0$ for a sufficiently small $h_S$ a first order) phase transition. 
In the $h_S\to 0$ limit, the behavior of the masses is more complicated. For instance, $m_{\eta^\prime}^2$ vanishes only in the complete absence of the anomaly, while the second order phase transition with $\phi_S\to0$ at $T\to T_\chi$ is present even in the case of including a $\propto \xi_2$ but no other term. In the presence of the anomaly, there exists a $h_S^{\star}(h_N)<0$ line, where $m_{\eta^\prime}^2=0$, which also depends on the anomaly terms and the parameterization. Below this $h_S\left(h_N\right)$ line, the model has to be extended with $\eta_N$ and $\eta_S$ condensates, which leads to a spontaneous CP breaking. 

We note that due to the wavefunction renormalization factors the pseudoscalar masses depend on the $\delta_N$ and $\delta_S$ explicit breaking parameter for the (axial) vectors and so does the Columbia plot if it is depicted as a function of the pion and kaon mass. However, this effect is very small and results in no qualitative modification. Nevertheless, for completeness $\delta_N$ and $\delta_S$ are rescaled as $h_N$ and $h_S$, respectively.

\bibliography{columbia}

\end{document}